\documentclass[11pt,a4paper]{article}
\usepackage{jcappub}
\usepackage{graphicx}
\usepackage{dcolumn}   
\usepackage{bm}        
\usepackage[dvipsnames]{xcolor}
\usepackage{amssymb, amsmath,framed,mathtools}

\expandafter\ifx\csname package@font\endcsname\relax\else
 \expandafter\expandafter
 \expandafter\usepackage
 \expandafter\expandafter
 \expandafter{\csname package@font\endcsname}
\fi
\def\bq{\begin{equation}}
\def\eq{\end{equation}}
\def\bqy{\begin{eqnarray}}
\def\eqy{\end{eqnarray}}

\def\p{\partial}

\def\p{\partial}

\def\cale{\mathcal{E}}
\def\calf{\mathcal{F}}

 \title{Is Life Most Likely Around Sun-like Stars?}
\author[a,b]{Manasvi Lingam}
\author[a,b]{Abraham Loeb}
\affiliation[a]{Harvard-Smithsonian Center for Astrophysics,
60 Garden Street, Cambridge, MA 02138, USA}
\affiliation[b]{Institute for Theory and Computation, Harvard University, Cambridge, MA 02138, USA}
\emailAdd{manasvi.lingam@cfa.harvard.edu}
\emailAdd{aloeb@cfa.harvard.edu}
\abstract{We consider the habitability of Earth-analogs around stars of different masses, which is regulated by the stellar lifetime, stellar wind-induced atmospheric erosion, and biologically active ultraviolet (UV) irradiance. By estimating the timescales associated with each of these processes, we show that they collectively impose limits on the habitability of Earth-analogs. We conclude that planets orbiting most M-dwarfs are not likely to host life, and that the highest probability of complex biospheres is for planets around K- and G-type stars. Our analysis suggests that the current existence of life near the Sun is slightly unusual, but not significantly anomalous.}

\begin{document}

\maketitle

\flushbottom

\section{Introduction} \label{SecIntro}
The discovery of thousands of exoplanets over the past decade has been accompanied by notable advances in our understanding of the many factors that are responsible for making a planet habitable \citep{Kal17}. Studies of habitability should attempt to find the right balance between complexity and transparency, while also being expressible in terms of basic physical parameters that can be deduced from observations. We will adopt this approach for studying how multiple stellar properties can regulate the habitability of planets in the habitable zone (HZ), i.e. the region around the host star where liquid water can exist on the planetary surface \citep{KWR93}. 

The first factor we take into account is the stellar lifetime, since it constitutes an upper bound on the timescale over which life-as-we-know-it can exist. The second phenomenon that we consider is the role of stellar winds in driving the erosion of planetary atmospheres because it is an effect that is particularly important for low-mass stars \citep{DLMC,LiLo17}. Our third consideration is the UV radiation environment, which has been suggested to have played an important role in facilitating prebiotic chemistry \citep{RBD14}, and the rise in oxygen levels \citep{CZM01} on Earth and other planets in the HZ.

In this paper, we will study the timescales associated with each of these processes, and how they collectively set fairly stringent limits on the habitability of Earth-like planets around stars of different masses. Our work has implications for the anthropic argument \citep{BT86} and its cosmological implications, since it addresses the question of why we find ourselves in a cosmic epoch where $\Omega_m \sim \Omega_\Lambda$ \citep{LBS16}. However, before embarking on our analysis, we caution the reader that we only evaluate a subset of potentially necessary (but not sufficient) conditions for life, and use heuristic order-of-magnitude estimates to illustrate conceptual points.

\section{Bioactive ultraviolet radiation and abiogenesis}

\begin{figure*}
$$
\begin{array}{cc}
  \includegraphics[width=7.2cm]{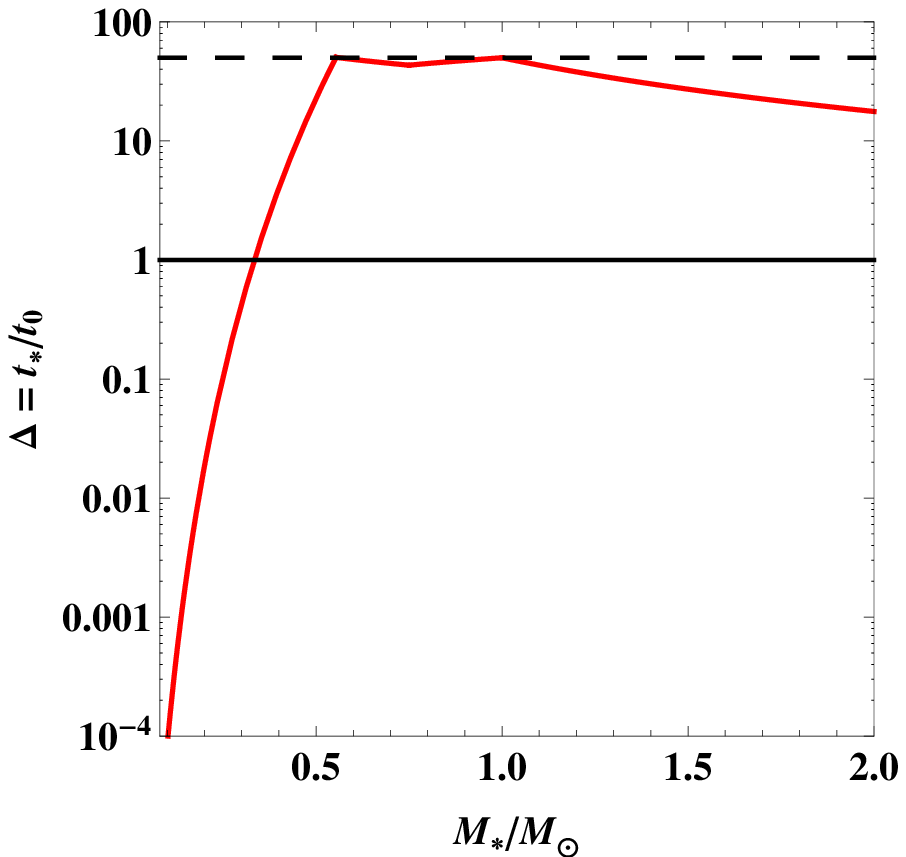} &  \includegraphics[width=7.1cm]{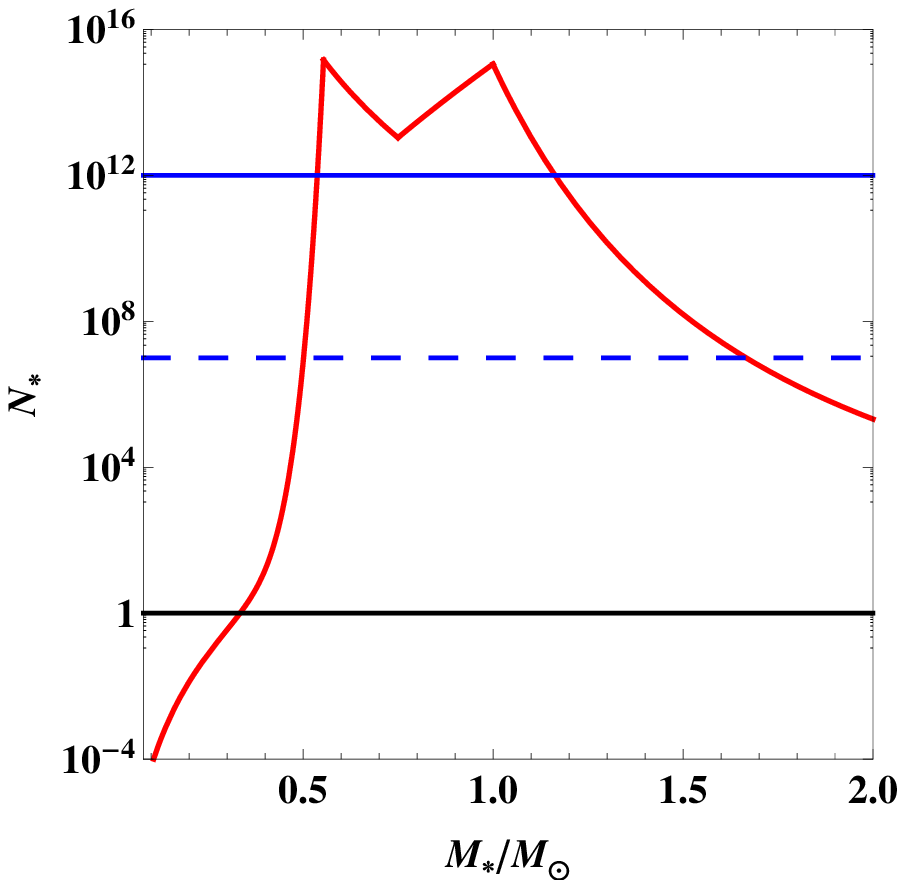}\\
\end{array}
$$
\caption{Left panel: maximum amplification factor $\Delta = t_\star/t_0$ as a function of stellar mass (in units of $M_\odot$). The dashed line denotes the solar value, while the solid line is the boundary below which abiogenesis would not occur. Right panel: peak diversity $\left(N_\star\right)$ attainable as a function of stellar mass. The blue solid and dashed lines represent the \emph{current} microbial and eukaryotic species diversity on Earth. The black solid line denotes the limit below which abiogenesis is not feasible on Earth-analogs.}
\label{Fig1}
\end{figure*}

A great deal remains unknown about the origin of life on Earth, such as the sites of abiogenesis and prebiotic systems chemistry \citep{Lu16}. With regards to the latter scenario, the role of ultraviolet (UV) radiation has been widely explored. UV light may play an important role in RNA polymerization \citep{MCG03}, resolving the asphaltization problem  \citep{BKC12} and enabling the synthesis of: (i)  pyrimidine ribonucleotides \citep{PGS09}, (ii) simple sugars \citep{RS12}, and (iii) the precursors of amino acids, nucleic acids, lipids and carbohydrates \citep{Pat15}. Arguments in favor of UV-mediated prebiotic pathways have been partly based on the above data, as well as the stability of RNA nucleotides to UV irradiation, suggesting that they might have arisen in a UV-rich environment \citep{Beck16}.\footnote{However, other energy sources may power prebiotic chemical reactions \citep{CS92}, and certain environments posited for the origin of life (e.g. hydrothermal vents) do not depend upon the availability of UV radiation \citep{BS10}.}

We shall posit henceforth that UV-driven prebiotic synthesis led to the origin of life, and that UV photochemistry constitutes the rate-limiting step in abiogenesis. We will consider an ``Earth-analog'' in the rest of the paper; by this term, we refer to an exoplanet whose basic physical parameters, e.g. the radius, effective temperature and surface pressure, are similar to those of Earth. The next step is to determine the biologically active (bioactive) UV irradiance $\mathcal{F}$ incident upon the Earth analog.\footnote{Since we have considered an Earth-analog, we have ignored the effects of atmospheric composition (e.g. hazes \citep{HS15}) on the amount of UV radiation reaching the surface. Although this factor is important, it is also hard to quantify and most studies concerning UV constraints prebiotic chemistry assume either Earth- or Mars-like atmospheres \citep{Ran17}.} The corresponding range of wavelengths is $200\,\mathrm{nm} \lesssim \lambda \lesssim 400\,\mathrm{nm}$, since photons with wavelengths $\lesssim 200$ nm are absorbed by CO$_2$ and H$_2$O in the atmosphere \citep{RS17}. To zeroth order, it can be assumed that, under the above assumptions, $\mathcal{F}$ depends only on the stellar mass $M_\star$. In reality, it is dependent on several other parameters such as the stellar age, rotation and activity \citep{SB14}. Using the data tabulated in Table 3 of Ref. \cite{RS15} for the biological UV flux at the surface, we adopt the following heuristic scaling relation:
\begin{eqnarray} \label{UVFlux}
    && \calf \sim \calf_\odot \left(\frac{M_\star}{M_\odot}\right)^3 \quad M_\star \lesssim M_\odot  \nonumber \\
    && \calf \sim \calf_\odot \left(\frac{M_\star}{M_\odot}\right) \quad \,\, M_\star \gtrsim M_\odot
\end{eqnarray}
which is valid in our region of interest ($M_\star \lesssim 2M_\odot$).\footnote{Our analysis can be further refined by introducing a more sophisticated piecewise power-law or polynomial ansatz for $\calf$, but the essential qualitative conclusions are mostly unchanged.} By assuming that the rates of prebiotic reactions necessary for abiogenesis are proportional to the bioactive UV flux and making use of (\ref{UVFlux}), we obtain
\begin{eqnarray} \label{AbioTS}
    && t_0 \sim t_\oplus \left(\frac{M_\star}{M_\odot}\right)^{-3} \quad M_\star \lesssim M_\odot  \nonumber \\
    && t_0 \sim t_\oplus \left(\frac{M_\star}{M_\odot}\right)^{-1} \quad M_\star \gtrsim M_\odot
\end{eqnarray}
where $t_0$ is the timescale for life to originate, and $t_\oplus \gtrsim 200$ Myr is a fiducial timescale for abiogenesis on Earth \citep{BBHM,Dod17,LL17}. Hence, for planets orbiting M-dwarfs, the timescale for the origin of life would be $\sim 10$ Gyr if the associated timescale on Earth is $ \sim 100$ Myr, in agreement with Ref. \cite{RWS17}. It would therefore appear logical to conclude that abiogenesis on M-dwarfs can occur, but it takes a long time.

In addition, one must also take into account the fact that planets around M-dwarfs are subject to rapid atmospheric erosion by the stellar wind \citep{LiLo17}. For low-mass M-dwarfs, the erosion timescale is of order $\sim 10-100$ Myr \citep{Aira17,DLMC,GG17,DJL18}. By employing constraints set by atmospheric erosion due to the stellar wind for a weakly magnetized planet (whose timescale we denote by $t_{SW}$) and the stellar lifetime, the maximum timescale $t_\star$ over which speciation can occur was presented in Ref. \cite{LL17}:
\begin{eqnarray} \label{MaxTS}
&& t_\star \sim 1.00\,t_\odot\,\left(\frac{M_\star}{M_\odot}\right)^{-2.5} \quad M_\star > 0.75 M_\odot \\
&& t_\star \sim 0.76\, t_\odot\, \left(\frac{M_\star}{M_\odot}\right)^{-3.5} \quad 0.55 M_\odot < M_\star \leq 0.75 M_\odot \nonumber \\
&& t_\star \sim 100\, t_\odot\, \left(\frac{M_\star}{M_\odot}\right)^{4.8} \quad \quad 0.08 M_\odot < M_\star \leq 0.55 M_\odot \nonumber
\end{eqnarray}
Here, $t_\odot \sim 10$ Gyr denotes the total lifetime of the Sun. The available empirical evidence on Earth indicates that the total number of species (species richness) can be modelled approximately via an exponential function \citep{Russ95,Ben09}; naturally, this exponential amplification will not proceed \emph{ad infinitum}. Hence, the species richness can be modeled as
\begin{equation} \label{Ndef}
N\left(t\right) =  \exp\left(\frac{t}{\tau}\right) - 1,
\end{equation}
where $\tau$ is the associated e-folding timescale. Since $N\left(t_0\right) = 1$, we can use (\ref{Ndef}) to obtain $t_0/\tau = \ln 2$. We can use this relation to express (\ref{Ndef}) in terms of $t_0$ as follows:
\begin{equation}
N\left(t\right) = 2^{t/t_0} - 1.
\end{equation}
It is important to note that the exponential amplification (driven by natural mutations) kicks in once abiogenesis has occurred, i.e. for $t > t_0$. Based on the preceding discussion, the value of $t_0$ is constrained by the availability of bioactive UV flux, and is governed by (\ref{AbioTS}). We can now construct a `peak' biological diversity \citep{LL17} by assuming that this exponential amplification occurs until $t = t_\star$, with $t_\star$ given by (\ref{MaxTS}). Thus, we introduce
\begin{equation}
    N_\star \equiv N\left(t_\star\right) = 2^\Delta - 1,
\end{equation}
where $\Delta = t_\star/t_0$. Hence, it is convenient to interpret $\Delta$ as the (maximum) amplification factor for evolution, and only the abiogenesis timescale $t_0$ is dependent on the available bioactive UV flux. Fig. \ref{Fig1} shows the values of $\Delta$ and $N_\star$ as a function of stellar mass. Using the stellar mass function to compute the weighted value of $N_\star$ \citep{LBS16} does not alter the right panel of Fig. \ref{Fig1} significantly, since the exponential factor in (\ref{Ndef}) is the dominant contribution. The scenario where the e-folding timescale $\tau$ in (\ref{Ndef}) is held constant has been plotted in Fig. 1 of Ref. \cite{LL17}, and the resultant implications are discussed in Sec. 3 of that paper.

For Earth-analogs around stars with $M_\star \lesssim 0.3 M_\odot$, we find $N_\star < 1$. Hence, such planets may have minimal chances of possessing life since their atmospheres would be stripped prior to the (relatively slow) emergence of life. From the left panel of Fig. \ref{Fig1}, we see that the amplification factor $\Delta$ is nearly constant for $0.5 \lesssim M_\star \lesssim 2 M_\odot$. Although massive stars have a shorter lifetime, they also have a higher fraction of the emitted energy in the bioactive UV range, implying that UV-mediated prebiotic pathways could operate at faster rates. Hence, the lower lifetime is counteracted by a correspondingly shorter abiogenesis timescale, thus resulting in a near-constant value of $\Delta$. The right panel of Fig. \ref{Fig1} displays a double peaked structure, and the peak species richness occurs for $0.5 \lesssim M_\star \lesssim M_\odot$. Furthermore, $N_\star$ for $M_\star \sim M_\odot$ approximately attains the global maximum value.

The above results may collectively explain why we find ourselves around a K- or G-type star, and not in the HZ of an M-dwarf, despite the latter being more numerous and characterized by long stellar lifetimes \citep{LBS16}. When the consequences of extreme space weather events arising from large flares and superflares are taken into account, the likelihood of complex biospheres around K- and M-dwarfs will be further diminished \citep{Lingam}. If we choose a lower bound of $\sim 0.5 M_\odot$ for stars to host complex biospheres and utilize Fig. 4 of Ref.  \cite{LBS16}, we find that terrestrial life at the present cosmic time has a probability of $\sim 10\%$. Clearly, our presence is far less anomalous compared to the scenario where all stars host complex life since that has a $0.1\%$ probability \citep{LBS16}.\footnote{Our Sun's Galactic orbit has a relatively low eccentricity \citep{RL08}, which can further boost its chances of habitability.}

Before proceeding further, we wish to point out a few caveats regarding the model. As noted earlier, stellar parameters other than $M_\star$ (especially the planetary system's age) will play a notable role in regulating $\calf$. Habitable planets around low-mass stars are capable of building up abiotic O$_2$ atmospheres (even up to $\sim 100$ bars) through several mechanisms \citep{WH14,LB15,HSS15,Mea17}. Given that more massive atmospheres will take longer to be eroded, there may exist a longer time interval for life to originate. However, elevated levels of ozone (formed via UV photolysis of O$_2$) would serve as a shield and prevent bioactive UV radiation from reaching the surface. Since there exist two opposing factors, it is unclear as to whether thick O$_2$ atmospheres would lead to beneficial or harmful ramifications.

Although we have not considered the role of flares thus far, they also give rise to both positive and negative consequences. On the one hand, it is plausible that flares could transiently deliver the requisite levels of UV radiation and Solar Energetic Particles (SEPs) to power prebiotic synthesis \citep{RWS17,LD18}. On the other hand, large flares on active stars are typically accompanied by high-fluence solar proton events and elevated levels of ionizing radiation, which may engender significant damage to the biosphere \citep{Dart11,Lingam}. Moreover, during these events, atmospheric losses are enhanced by $1$-$2$ orders of magnitude \citep{DHL}.

Lastly, we note that (\ref{AbioTS}) implies that $t_0$ varies by 2-3 orders of magnitudes in the transition from M- to G-type stars; see Refs. \cite{LD02} and \cite{ST12} for differing Bayesian analyses of $t_0$ and the ensuing implications. Furthermore, Fig. \ref{Fig1} indicates that stars $\lesssim 0.3 M_\odot$ are unlikely to host inhabited planets. These trends jointly suggest that the fraction of life-bearing planets might be rather low in the current epoch of the Universe.

\section{Ultraviolet radiation and complex life}
\begin{table*}
\begin{minipage}{126mm}
\caption{Characteristic timescales for Earth-analogs orbiting stars of different masses}
\label{Tab1}
\begin{tabular}{|c|c|c|c|c|c|}
\hline 
Star mass ($M_\odot$) & Stellar lifetime (yr) & $t_{SW}$ (yr) & $t_0$ (yr) & $t_{O_2}$ (yr) & $t_\ell$ (yr)\tabularnewline
\hline 
\hline 
$0.1$ & $6.7 \times 10^{12}$ & $1.7 \times 10^7$ & $2 \times 10^{11}$ & $1.1 \times 10^7$ & $2 \times 10^{11}$\tabularnewline
\hline 
$0.3$ & $5.1 \times 10^{11}$ & $3.2 \times 10^9$ & $7.4 \times 10^9$ & $1.4 \times 10^8$ & $7.5 \times 10^9$\tabularnewline
\hline 
$0.6$ & $4.5 \times 10^{10}$ & $8.8 \times 10^{10}$ & $9.3 \times 10^8$ & $6.8 \times 10^8$ & $1.6 \times 10^9$\tabularnewline
\hline 
$1.0$ & $10^{10}$ & $10^{12}$ & $2 \times 10^8$ & $2.2 \times 10^9$ & $2.4 \times 10^9$\tabularnewline
\hline 
$1.5$ & $3.6 \times 10^{9}$ & $6.9 \times 10^{12}$ & $1.3 \times 10^8$ & $5.8 \times 10^8$ & $7.1 \times 10^8$\tabularnewline
\hline 
\end{tabular}
\medskip
\end{minipage}
\end{table*}

\begin{figure}
\quad\quad\quad \includegraphics[width=7.2cm]{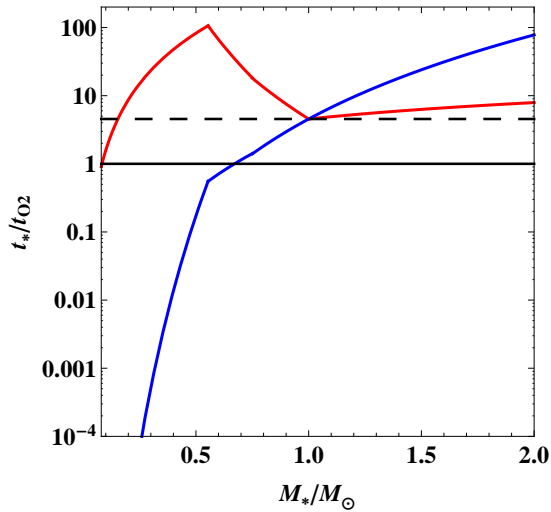} \\
\caption{The ratio of the timescale for oxygenation ($t_{O_2}$) and the maximum duration over which evolution can take place ($t_\star$). The dashed line denotes the solar value, while the unbroken line is the boundary below which sufficient levels of oxygenation would not be feasible. The red and blue curves include and exclude the emission from stellar chromospheric activity, respectively.}
\label{Fig2}
\end{figure}

Fig. \ref{Fig1} implies that stars with $M_\star \lesssim 0.3 M_\odot$ may not host life-bearing planets since atmospheric erosion occurs faster than abiogenesis. Although bioactive UV radiation can power prebiotic chemistry, its propensity for causing DNA damage is well-known. Several studies have focused on this aspect \citep{Cock99,SK03,Kal17} since the absence of an ozone layer leads to higher levels of biologically effective UV radiation reaching the surface \citep{Dart11}. Typically, the UV flux also increases with the stellar mass and decreases with age.

Here, we shall explore a different role that UV radiation could play in facilitating the rise of complex life. On Earth, the following events are believed to have occurred in roughly chronological order: (i) oxygenic photosynthesis, (ii) Great Oxygenation Event (GOE), (iii) origin of  eukaryotes, (iv) Neoproterozoic Oxygenation Event (NOE), and (v) emergence of large and complex organisms \citep{Can05,HB11,LRP14,KN17}. The causes and timing of all these events are subject to large uncertainties; consequently, it is unclear as to whether the increase in oxygen levels had a cause-or-effect relationship with life. Despite this variability, it appears plausible that oxygen is a necessary requirement for complex life on habitable planets and thus constitutes a vital limiting step \citep{CG05,AHK17,Manasvi}. 

One of the main hypotheses for the GOE suggests that biogenic methane (produced by methanogens) undergoes UV photolysis, thereby leading to hydrogen escape to escape and irreversible oxidation \citep{CZM01,CC05}. It has also been suggested that the buildup of oxygen on Earth may have occurred through the photolysis of water by UV radiation \citep{Carv81}. On low-mass stars, UV photolysis of CO$_2$ also leads to significant production of abiotic oxygen \citep{TLMV14}. In these scenarios, it is clear that UV light plays a significant role in enabling the rise of atmospheric oxygen.\footnote{On the other hand, if the GOE did not occur either due to biotic or abiotic UV photolysis, expressing the timescale for oxygenation in terms of basic stellar parameters is not feasible since the actual causes would be related to planetary processes.} In addition to the putative role of UV irradiance (determined by stellar physics), many complex biogeochemical governing factors \citep{KS02} are not considered herein. 

A similar line of reasoning concerning the role of UV photolysis in raising oxygen levels was adopted in Ref. \cite{Liv99} to arrive at the characteristic timescale for life. Here, we shall focus on quantifying $t_{O_2}$, i.e. the timescale associated with initial oxygenation under the assumption that UV photolysis played a significant role. Since a significant fraction ($\gtrsim 50\%$) of both water and methane photodissociation occurs via Ly$\alpha$, we will consider the Ly$\alpha$ flux to be a proxy for the rate of oxygenation. Thus, in analogy to (\ref{UVFlux}) and (\ref{AbioTS}), we identify the following scaling relations:
\begin{eqnarray} \label{OxyTime}
    && t_{O_2} \sim 0.22\, t_\odot \left(\frac{M_\star}{M_\odot}\right)^{2.3} \quad\quad M_\star \lesssim M_\odot  \nonumber \\
    && t_{O_2} \sim 0.22\, t_\odot \left(\frac{M_\star}{M_\odot}\right)^{-3.3} \quad\, M_\star \gtrsim M_\odot
\end{eqnarray}
For $M_\star \lesssim M_\odot$, we have used Fig. 9 and Table 5 of Ref.  \cite{LFA13} for $P > 25$ days along with the stellar effective temperature-mass relation, $T_\mathrm{eff} \propto M_\star^{0.5}$. On the other hand, the sample size for $M_\star \gtrsim M_\odot$ is relatively small. For F-type stars, we have utilized Ref. \cite{LS93}, Table 4 of Ref. \cite{LFA13}, and Table 6 of Ref. \cite{RS15}. We reiterate that, for the sake of simplicity, we have not taken into account the effects of stellar rotation rate on the emitted Ly$\alpha$ flux. The presence of a positive exponent for $M_\star \lesssim M_\odot$ may seem surprising at first glimpse, since low-mass stars would be expected to radiate fewer far-UV (FUV) photons. In turn, this should lower the FUV flux, and thereby increase the associated timescale, leading to a negative exponent; instead, we see from (\ref{OxyTime}) that the exponent is positive for $M_\star \lesssim M_\odot$. The reason stems from the fact that low-mass stars are known to be characterized by significant UV emission from the chromospheric and transition regions that lead to elevated FUV fluxes \citep{TLMV14,LFF14,SB14}. 

We have also relied upon the fact that the GOE occurred on Earth $\sim 2.4$ Gyr ago \citep{LRP14}.\footnote{We will not analyze the NOE herein despite its significance from an evolutionary perspective, since its timing and causal relationship with biota remains unclear \citep{OS12}.} Note that $t_{O_2} \lesssim t_\cale$, where $t_\cale$ is the timescale for eukaryogenesis \citep{PLK11}, implying that $t_{O_2}$ can be viewed as a potential lower bound for the emergence of ``complex'' life. We emphasize that (\ref{OxyTime}) has been constructed based on observational evidence and therefore includes the Ly$\alpha$ flux arising from stellar chromospheres and transition regions. In contrast, if their role is ignored, we may use \citep{Liv99}:
\begin{equation}\label{OxyTimeNF}
    t_{O_2} \sim 0.22\, t_\odot \left(\frac{M_\star}{M_\odot}\right)^{-6.6}.
\end{equation}
A striking difference between (\ref{OxyTime}) and (\ref{OxyTimeNF}) is that, for $M_\star \lesssim M_\odot$, the former relation has a positive spectral index while the latter has a negative value. Hence, in the presence (absence) of chromospheric activity, sufficient levels of oxygenation for complex life to flourish would arise over faster (slower) timescales for low-mass stars.

We are therefore in a position to compare $t_{O_2}$ with $t_\star$. If $t_\star > t_{O_2}$, the Earth-analog would be potentially capable of giving rise to complex life, and vice-versa. In Fig. \ref{Fig2}, the timescales for oxygenation, with and without the contributions from stellar chromospheres and transition regions - (\ref{OxyTime}) and (\ref{OxyTimeNF}) respectively - have been compared to $t_\star$. When (\ref{OxyTime}) is taken into account, levels of oxygen comparable to the GOE are always attainable. In contrast, when (\ref{OxyTimeNF}) is used, only stars with $M_\star \gtrsim 0.67 M_\odot$ can fulfill this criterion.

One may also consider the timescale, $t_\ell = t_0 + t_{O_2}$, because $t_0$ and $t_{O_2}$ are approximately the timescales for ``simple'' and ``complex'' life respectively. Hence, their sum ($t_\ell$) is a heuristic measure of the total time required for complex life to originate, based on the assumption that UV radiation plays an important role in regulating both $t_0$ and $t_{O_2}$. From (\ref{AbioTS}) and (\ref{OxyTime}), we find that the branch $M_\star \lesssim M_\odot$ is characterized by non-monotonic behaviour. Solving for $\p t_\ell/\p M_\star = 0$, we find that a minimum value is attained at $M_\star \approx 0.67 M_\odot$ corresponding to $t_\ell \sim 1.5$ Gyr.\footnote{For the Earth-Sun system, we obtain $t_\ell \sim 2.4$ Gyr, slightly lower than the eukaryogenesis timescale of $\sim 2.6$ Gyr.} Our analysis implies, \emph{ceteris paribus}, that the fastest time for the emergence of complex life would occur on a K-type star when we restrict ourselves to stars with $M_\star \lesssim M_\odot$. If we consider the branch $M_\star \gtrsim M_\odot$, we find that $t_\ell$ decreases monotonically with the mass. In this case, however, the shorter stellar lifetime would imply that speciation would take place over a shorter timescale, as seen from the right panel of Fig. \ref{Fig1}. 

We conclude by observing that (\ref{OxyTime}) indicates that the timescale for raising oxygen (by means of UV photolysis) to levels that would enable complex life to emerge is $\sim 100$ times lower on Earth-analogs orbiting M-dwarfs. This might imply that intelligent life could also arise much faster on certain planets. A similar result was presented in Ref. \cite{McK96}, albeit based on the analysis of planetary (and not stellar) constraints. However, it must be emphasized that rapid oxygenation could also result in detrimental consequences, since O$_2$ was toxic to early organisms on Earth.

\section{Conclusion}
In this paper, we explored multiple constraints on the potential habitability of Earth-analogs orbiting stars of different masses. We chose the stellar mass as our governing parameter, since it can be measured easily relative to other parameters (e.g. stellar age and activity), and it enables us to identify appropriate stellar systems for future searches of extraterrestrial life. Naturally, there exist several other stellar and planetary parameters which also play a significant role in UV habitability that were not considered here. We considered constraints set by: (i) atmospheric erosion driven by the stellar wind, (ii) stellar lifetime, (iii) availability of UV radiation to power prebiotic chemical reactions, and (iv) UV photolysis of molecules to enable oxygenation. The characteristic timescales for all these processes are summarized in Table \ref{Tab1} for Earth-analogs around different stars.

We found that stars $\lesssim 0.3 M_\odot$ are unlikely to host life-bearing planets because their atmospheres would be eroded prior to abiogenesis. Hence, the probability that the planets in the HZ of Proxima Centauri, TRAPPIST-1 and LHS 1140 are inhabited is low. Planets around higher-mass M-dwarfs could potentially have life but their prospects of hosting Earth-like complex biospheres are not high. We concluded that K- and G-type stars are most likely to host complex biospheres, as seen from the plot of maximal species richness (number of species) in Fig. \ref{Fig1}. In fact, their peak species richness is comparable since K-dwarfs are longer lived than G-type stars but, on the other hand, the associated timescale for abiogenesis is also longer. Based on criteria (iii) and (iv), we found that Earth-analogs around stars with $M_\star \approx 0.67 M_\odot$ may take the least amount of time for complex (eukaryotic-type) life to originate. Planets in the HZ of F-type stars could, due to elevated levels of UV radiation, achieve abiogenesis and oxygenation more rapidly compared to the Earth.\footnote{On the other hand, the elevated UV levels can prove to be deleterious to the evolution and sustenance of complex life.} Conversely, as these stars have a shorter lifetime, their planets' peak biodiversity might be lower when compared to the Earth-Sun system.

Thus, based on these considerations, it seems reasonable to suggest that future searches for life should prioritize K- and G-type stars insofar our analysis based on (i)-(iv) is concerned, although early M-dwarfs and F-type stars may also represent promising targets; a similar position was advocated in Refs. \cite{KWR93,HA14,LL17,LiL18,LL18}. Lastly, our analysis also suggests that our existence around a Sun-like star in the present epoch is not particularly anomalous \citep{LBS16}.

\acknowledgments
We thank Charlie Conroy and Sukrit Ranjan for the helpful comments, and our reviewer for the constructive report. This work was supported by the Breakthrough Prize Foundation for the Starshot Initiative, Harvard University's Faculty of Arts and Sciences, and the Institute for Theory and Computation (ITC) at Harvard University. 


\begin{thebibliography}{10}

\bibitem{Kal17}
L.~{Kaltenegger}, \emph{{How to Characterize Habitable Worlds and Signs of
  Life}},
  \href{https://doi.org/10.1146/annurev-astro-082214-122238}{\emph{Annu. Rev.
  Astron. Astrophys.} {\bfseries 55} (2017) 433--485}.

\bibitem{KWR93}
J.~F. {Kasting}, D.~P. {Whitmire} and R.~T. {Reynolds}, \emph{{Habitable Zones
  around Main Sequence Stars}},
  \href{https://doi.org/10.1006/icar.1993.1010}{\emph{Icarus} {\bfseries 101}
  (1993) 108--128}.

\bibitem{DLMC}
C.~{Dong}, M.~{Lingam}, Y.~{Ma} and O.~{Cohen}, \emph{{Is Proxima Centauri b
  Habitable? A Study of Atmospheric Loss}},
  \href{https://doi.org/10.3847/2041-8213/aa6438}{\emph{Astrophys. J. Lett.}
  {\bfseries 837} (2017) L26},
  [\href{https://arxiv.org/abs/1702.04089}{{\ttfamily 1702.04089}}].

\bibitem{LiLo17}
M.~{Lingam} and A.~{Loeb}, \emph{{Physical constraints on the likelihood of
  life on exoplanets}},
  \href{https://doi.org/10.1017/S1473550417000179}{\emph{Int. J. Astrobiol.}
  {\bfseries 17} (2018) 116--126},
  [\href{https://arxiv.org/abs/1707.02996}{{\ttfamily 1707.02996}}].

\bibitem{RBD14}
K.~{Ruiz-Mirazo}, C.~{Briones} and A.~{de la Escosura}, \emph{{Prebiotic
  Systems Chemistry: New Perspectives for the Origins of Life}},
  \href{https://doi.org/10.1021/cr2004844}{\emph{Chem. Rev.} {\bfseries 114}
  (2014) 285--366}.

\bibitem{CZM01}
D.~C. {Catling}, K.~J. {Zahnle} and C.~P. {McKay}, \emph{{Biogenic Methane,
  Hydrogen Escape, and the Irreversible Oxidation of Early Earth}},
  \href{https://doi.org/10.1126/science.1061976}{\emph{Science} {\bfseries 293}
  (2001) 839--843}.

\bibitem{BT86}
J.~D. {Barrow} and F.~J. {Tipler}, \emph{{The Anthropic Cosmological
  Principle}}.
\newblock Oxford: Clarendon Press, 1986.

\bibitem{LBS16}
A.~{Loeb}, R.~A. {Batista} and D.~{Sloan}, \emph{{Relative likelihood for life
  as a function of cosmic time}},
  \href{https://doi.org/10.1088/1475-7516/2016/08/040}{\emph{J. Cosmol.
  Astropart. Phys.} {\bfseries 8} (2016) 040},
  [\href{https://arxiv.org/abs/1606.08448}{{\ttfamily 1606.08448}}].

\bibitem{Lu16}
P.~L. {Luisi}, \emph{The Emergence of Life: From Chemical Origins to Synthetic
  Biology}.
\newblock Cambridge Univ. Press, 2016.

\bibitem{MCG03}
A.~Y. {Mulkidjanian}, D.~A. {Cherepanov} and M.~Y. {Galperin}, \emph{{Survival
  of the fittest before the beginning of life: selection of the first
  oligonucleotide-like polymers by UV light}},
  \href{https://doi.org/10.1186/1471-2148-3-12}{\emph{BMC Evol. Biol.}
  {\bfseries 3} (2003) 12}.

\bibitem{BKC12}
S.~A. {Benner}, H.-J. {Kim} and M.~A. {Carrigan}, \emph{{Asphalt, Water, and
  the Prebiotic Synthesis of Ribose, Ribonucleosides, and RNA}},
  \href{https://doi.org/10.1021/ar200332w}{\emph{Acc. Chem. Res.} {\bfseries
  45} (2012) 2025--2034}.

\bibitem{PGS09}
M.~W. {Powner}, B.~{Gerland} and J.~D. {Sutherland}, \emph{{Synthesis of
  activated pyrimidine ribonucleotides in prebiotically plausible conditions}},
  \href{https://doi.org/10.1038/nature08013}{\emph{Nature} {\bfseries 459}
  (2009) 239--242}.

\bibitem{RS12}
D.~{Ritson} and J.~D. {Sutherland}, \emph{{Prebiotic synthesis of simple sugars
  by photoredox systems chemistry}},
  \href{https://doi.org/10.1038/nchem.1467}{\emph{Nat. Chem.} {\bfseries 4}
  (2012) 895--899}.

\bibitem{Pat15}
B.~H. {Patel}, C.~{Percivalle}, D.~J. {Ritson}, C.~D. {Duffy} and J.~D.
  {Sutherland}, \emph{{Common origins of RNA, protein and lipid precursors in a
  cyanosulfidic protometabolism}},
  \href{https://doi.org/10.1038/nchem.2202}{\emph{Nat. Chem.} {\bfseries 7}
  (2015) 301--307}.

\bibitem{Beck16}
A.~A. {Beckstead}, Y.~{Zhang}, M.~S. {de Vries} and B.~{Kohler}, \emph{{Life in
  the light: nucleic acid photoproperties as a legacy of chemical evolution}},
  \href{https://doi.org/10.1039/C6CP04230A}{\emph{Phys. Chem. Chem. Phys.}
  {\bfseries 18} (2016) 24228--24238}.

\bibitem{CS92}
C.~{Chyba} and C.~{Sagan}, \emph{{Endogenous production, exogenous delivery and
  impact-shock synthesis of organic molecules: an inventory for the origins of
  life}}, \href{https://doi.org/10.1038/355125a0}{\emph{Nature} {\bfseries 355}
  (1992) 125--132}.

\bibitem{BS10}
I.~{Budin} and J.~W. {Szostak}, \emph{{Expanding Roles for Diverse Physical
  Phenomena During the Origin of Life}},
  \href{https://doi.org/10.1146/annurev.biophys.050708.133753}{\emph{Annu. Rev.
  Biophys.} {\bfseries 39} (2010) 245--263}.

\bibitem{HS15}
K.~{Heng} and A.~P. {Showman}, \emph{{Atmospheric Dynamics of Hot Exoplanets}},
  \href{https://doi.org/10.1146/annurev-earth-060614-105146}{\emph{Annu. Review
  Earth Planet. Sci.} {\bfseries 43} (2015) 509--540},
  [\href{https://arxiv.org/abs/1407.4150}{{\ttfamily 1407.4150}}].

\bibitem{Ran17}
S.~{Ranjan}, R.~{Wordsworth} and D.~D. {Sasselov}, \emph{{Atmospheric
  Constraints on the Surface UV Environment of Mars at 3.9 Ga Relevant to
  Prebiotic Chemistry}},
  \href{https://doi.org/10.1089/ast.2016.1596}{\emph{Astrobiology} {\bfseries
  17} (2017) 687--708}, [\href{https://arxiv.org/abs/1701.01373}{{\ttfamily
  1701.01373}}].

\bibitem{RS17}
S.~{Ranjan} and D.~D. {Sasselov}, \emph{{Constraints on the Early Terrestrial
  Surface UV Environment Relevant to Prebiotic Chemistry}},
  \href{https://doi.org/10.1089/ast.2016.1519}{\emph{Astrobiology} {\bfseries
  17} (2017) 169--204}, [\href{https://arxiv.org/abs/1610.06223}{{\ttfamily
  1610.06223}}].

\bibitem{SB14}
E.~L. {Shkolnik} and T.~S. {Barman}, \emph{{HAZMAT. I. The Evolution of Far-UV
  and Near-UV Emission from Early M Stars}},
  \href{https://doi.org/10.1088/0004-6256/148/4/64}{\emph{Astron. J.}
  {\bfseries 148} (2014) 64},
  [\href{https://arxiv.org/abs/1407.1344}{{\ttfamily 1407.1344}}].

\bibitem{RS15}
S.~{Rugheimer}, A.~{Segura}, L.~{Kaltenegger} and D.~{Sasselov}, \emph{{UV
  Surface Environment of Earth-like Planets Orbiting FGKM Stars through
  Geological Evolution}},
  \href{https://doi.org/10.1088/0004-637X/806/1/137}{\emph{Astrophys. J.}
  {\bfseries 806} (2015) 137},
  [\href{https://arxiv.org/abs/1506.07200}{{\ttfamily 1506.07200}}].

\bibitem{BBHM}
E.~A. {Bell}, P.~{Boehnke}, T.~M. {Harrison} and W.~L. {Mao},
  \emph{{Potentially biogenic carbon preserved in a 4.1 billion-year-old
  zircon}}, \href{https://doi.org/10.1073/pnas.1517557112}{\emph{Proc. Natl.
  Acad. Sci. USA} {\bfseries 112} (2015) 14518--14521}.

\bibitem{Dod17}
M.~S. {Dodd}, D.~{Papineau}, T.~{Grenne}, J.~F. {Slack}, M.~{Rittner},
  F.~{Pirajno} et~al., \emph{{Evidence for early life in Earth's oldest
  hydrothermal vent precipitates}},
  \href{https://doi.org/10.1038/nature21377}{\emph{Nature} {\bfseries 543}
  (2017) 60--64}.

\bibitem{LL17}
M.~{Lingam} and A.~{Loeb}, \emph{{Reduced Diversity of Life around Proxima
  Centauri and TRAPPIST-1}},
  \href{https://doi.org/10.3847/2041-8213/aa8860}{\emph{Astrophys. J. Lett.}
  {\bfseries 846} (2017) L21},
  [\href{https://arxiv.org/abs/1707.07007}{{\ttfamily 1707.07007}}].

\bibitem{RWS17}
S.~{Ranjan}, R.~{Wordsworth} and D.~D. {Sasselov}, \emph{{The Surface UV
  Environment on Planets Orbiting M Dwarfs: Implications for Prebiotic
  Chemistry and the Need for Experimental Follow-up}},
  \href{https://doi.org/10.3847/1538-4357/aa773e}{\emph{Astrophys. J.}
  {\bfseries 843} (2017) 110},
  [\href{https://arxiv.org/abs/1705.02350}{{\ttfamily 1705.02350}}].

\bibitem{Aira17}
V.~S. {Airapetian}, A.~{Glocer}, G.~V. {Khazanov}, R.~O.~P. {Loyd},
  K.~{France}, J.~{Sojka} et~al., \emph{{How Hospitable Are Space Weather
  Affected Habitable Zones? The Role of Ion Escape}},
  \href{https://doi.org/10.3847/2041-8213/836/1/L3}{\emph{Astrophys. J. Lett.}
  {\bfseries 836} (2017) L3}.

\bibitem{GG17}
K.~{Garcia-Sage}, A.~{Glocer}, J.~J. {Drake}, G.~{Gronoff} and O.~{Cohen},
  \emph{{On the Magnetic Protection of the Atmosphere of Proxima Centauri b}},
  \href{https://doi.org/10.3847/2041-8213/aa7eca}{\emph{Astrophys. J. Lett.}
  {\bfseries 844} (2017) L13}.

\bibitem{DJL18}
C.~{Dong}, M.~{Jin}, M.~{Lingam}, V.~S. {Airapetian}, Y.~{Ma} and B.~{van der
  Holst}, \emph{{Atmospheric escape from the TRAPPIST-1 planets and
  implications for habitability}},
  \href{https://doi.org/10.1073/pnas.1708010115}{\emph{Proc. Natl. Acad. Sci.
  USA} {\bfseries 115} (2018) 260--265},
  [\href{https://arxiv.org/abs/1705.05535}{{\ttfamily 1705.05535}}].

\bibitem{Russ95}
D.~{Russell}, \emph{{Biodiversity and Time Scales for the Evolution of
  Extraterrestrial Intelligence}},  in \emph{Progress in the Search for
  Extraterrestrial Life.} (G.~S. {Shostak}, ed.), vol.~74 of \emph{Astronomical
  Society of the Pacific Conference Series}, pp.~143--151, Astronomical Society
  of the Pacific, 1995.

\bibitem{Ben09}
M.~J. {Benton}, \emph{{The Red Queen and the Court Jester: Species Diversity
  and the Role of Biotic and Abiotic Factors Through Time}},
  \href{https://doi.org/10.1126/science.1157719}{\emph{Science} {\bfseries 323}
  (2009) 728}.

\bibitem{Lingam}
M.~{Lingam} and A.~{Loeb}, \emph{{Risks for life on habitable planets from
  superflares of their host stars}},
  \href{https://doi.org/10.3847/1538-4357/aa8e96}{\emph{Astrophys. J.}
  {\bfseries 848} (2017) 41},
  [\href{https://arxiv.org/abs/1708.04241}{{\ttfamily 1708.04241}}].

\bibitem{RL08}
J.~A. {Robles}, C.~H. {Lineweaver}, D.~{Grether}, C.~{Flynn}, C.~A. {Egan},
  M.~B. {Pracy} et~al., \emph{{A Comprehensive Comparison of the Sun to Other
  Stars: Searching for Self-Selection Effects}},
  \href{https://doi.org/10.1086/589985}{\emph{Astrophys. J.} {\bfseries 684}
  (2008) 691--706}, [\href{https://arxiv.org/abs/0805.2962}{{\ttfamily
  0805.2962}}].

\bibitem{WH14}
R.~{Wordsworth} and R.~{Pierrehumbert}, \emph{{Abiotic Oxygen-dominated
  Atmospheres on Terrestrial Habitable Zone Planets}},
  \href{https://doi.org/10.1088/2041-8205/785/2/L20}{\emph{Astrophys. J. Lett.}
  {\bfseries 785} (2014) L20},
  [\href{https://arxiv.org/abs/1403.2713}{{\ttfamily 1403.2713}}].

\bibitem{LB15}
R.~{Luger} and R.~{Barnes}, \emph{{Extreme Water Loss and Abiotic O2Buildup on
  Planets Throughout the Habitable Zones of M Dwarfs}},
  \href{https://doi.org/10.1089/ast.2014.1231}{\emph{Astrobiology} {\bfseries
  15} (2015) 119--143}, [\href{https://arxiv.org/abs/1411.7412}{{\ttfamily
  1411.7412}}].

\bibitem{HSS15}
C.~E. {Harman}, E.~W. {Schwieterman}, J.~C. {Schottelkotte} and J.~F.
  {Kasting}, \emph{{Abiotic O$_{2}$ Levels on Planets around F, G, K, and M
  Stars: Possible False Positives for Life?}},
  \href{https://doi.org/10.1088/0004-637X/812/2/137}{\emph{Astrophys. J.}
  {\bfseries 812} (2015) 137},
  [\href{https://arxiv.org/abs/1509.07863}{{\ttfamily 1509.07863}}].

\bibitem{Mea17}
V.~S. {Meadows}, \emph{{Reflections on O$_{2}$ as a Biosignature in
  Exoplanetary Atmospheres}},
  \href{https://doi.org/10.1089/ast.2016.1578}{\emph{Astrobiology} {\bfseries
  17} (2017) 1022--1052}.

\bibitem{LD18}
M.~{Lingam}, C.~{Dong}, X.~{Fang}, B.~M. {Jakosky} and A.~{Loeb}, \emph{{The
  Propitious Role of Solar Energetic Particles in the Origin of Life}},
  \href{https://doi.org/10.3847/1538-4357/aa9fef}{\emph{Astrophys. J.}
  {\bfseries 853} (2018) 10},
  [\href{https://arxiv.org/abs/1801.05781}{{\ttfamily 1801.05781}}].

\bibitem{Dart11}
L.~R. {Dartnell}, \emph{{Ionizing Radiation and Life}},
  \href{https://doi.org/10.1089/ast.2010.0528}{\emph{Astrobiology} {\bfseries
  11} (2011) 551--582}.

\bibitem{DHL}
C.~{Dong}, Z.~{Huang}, M.~{Lingam}, G.~{T{\'o}th}, T.~{Gombosi} and
  A.~{Bhattacharjee}, \emph{{The Dehydration of Water Worlds via Atmospheric
  Losses}}, \href{https://doi.org/10.3847/2041-8213/aa8a60}{\emph{Astrophys. J.
  Lett.} {\bfseries 847} (2017) L4},
  [\href{https://arxiv.org/abs/1709.01219}{{\ttfamily 1709.01219}}].

\bibitem{LD02}
C.~H. {Lineweaver} and T.~M. {Davis}, \emph{{Does the Rapid Appearance of Life
  on Earth Suggest that Life Is Common in the Universe?}},
  \href{https://doi.org/10.1089/153110702762027871}{\emph{Astrobiology}
  {\bfseries 2} (2002) 293--304},
  [\href{https://arxiv.org/abs/astro-ph/0205014}{{\ttfamily
  astro-ph/0205014}}].

\bibitem{ST12}
D.~S. {Spiegel} and E.~L. {Turner}, \emph{{Bayesian analysis of the
  astrobiological implications of life's early emergence on Earth}},
  \href{https://doi.org/10.1073/pnas.1111694108}{\emph{Proc. Natl. Acad. Sci.
  USA} {\bfseries 109} (2012) 395--400},
  [\href{https://arxiv.org/abs/1107.3835}{{\ttfamily 1107.3835}}].

\bibitem{Cock99}
C.~S. {Cockell}, \emph{{Carbon Biochemistry and the Ultraviolet Radiation
  Environments of F, G, and K Main Sequence Stars}},
  \href{https://doi.org/10.1006/icar.1999.6167}{\emph{Icarus} {\bfseries 141}
  (1999) 399--407}.

\bibitem{SK03}
A.~{Segura}, K.~{Krelove}, J.~F. {Kasting}, D.~{Sommerlatt}, V.~{Meadows},
  D.~{Crisp} et~al., \emph{{Ozone Concentrations and Ultraviolet Fluxes on
  Earth-Like Planets Around Other Stars}},
  \href{https://doi.org/10.1089/153110703322736024}{\emph{Astrobiology}
  {\bfseries 3} (2003) 689--708}.

\bibitem{Can05}
D.~E. {Canfield}, \emph{{The Early History of Atmospheric Oxygen: Homage to
  Robert M. Garrels}},
  \href{https://doi.org/10.1146/annurev.earth.33.092203.122711}{\emph{Annu.
  Rev. Earth Planet. Sci.} {\bfseries 33} (2005) 1--36}.

\bibitem{HB11}
M.~F. {Hohmann-Marriott} and R.~E. {Blankenship}, \emph{{Evolution of
  Photosynthesis}},
  \href{https://doi.org/10.1146/annurev-arplant-042110-103811}{\emph{Annu. Rev.
  Plant Biol.} {\bfseries 62} (2011) 515--548}.

\bibitem{LRP14}
T.~W. {Lyons}, C.~T. {Reinhard} and N.~J. {Planavsky}, \emph{{The rise of
  oxygen in Earth's early ocean and atmosphere}},
  \href{https://doi.org/10.1038/nature13068}{\emph{Nature} {\bfseries 506}
  (2014) 307--315}.

\bibitem{KN17}
A.~H. {Knoll} and M.~A. {Nowak}, \emph{{The timetable of evolution}},
  \href{https://doi.org/10.1126/sciadv.1603076}{\emph{Sci. Adv.} {\bfseries 3}
  (2017) e1603076}.

\bibitem{CG05}
D.~C. {Catling}, C.~R. {Glein}, K.~J. {Zahnle} and C.~P. {McKay}, \emph{{Why
  O$_{2}$ Is Required by Complex Life on Habitable Planets and the Concept of
  Planetary ``Oxygenation Time''}},
  \href{https://doi.org/10.1089/ast.2005.5.415}{\emph{Astrobiology} {\bfseries
  5} (2005) 415--438}.

\bibitem{AHK17}
A.~H. {Knoll}, \emph{{Biogeochemistry: Food for early animal evolution}},
  \href{https://doi.org/10.1038/nature23539}{\emph{Nature} {\bfseries 548}
  (2017) 528--530}.

\bibitem{Manasvi}
M.~{Lingam} and A.~{Loeb}, \emph{{Subsurface Exolife}},
  \href{https://doi.org/10.1017/S1473550418000083}{\emph{Int. J. Astrobiol.}
  (2018) }, [\href{https://arxiv.org/abs/1711.09908}{{\ttfamily 1711.09908}}].

\bibitem{CC05}
D.~C. {Catling} and M.~W. {Claire}, \emph{{How Earth's atmosphere evolved to an
  oxic state: A status report}},
  \href{https://doi.org/10.1016/j.epsl.2005.06.013}{\emph{Earth Planet. Sci.
  Lett.} {\bfseries 237} (2005) 1--20}.

\bibitem{Carv81}
J.~H. {Carver}, \emph{{Prebiotic atmospheric oxygen levels}},
  \href{https://doi.org/10.1038/292136a0}{\emph{Nature} {\bfseries 292} (1981)
  136--138}.

\bibitem{TLMV14}
F.~{Tian}, K.~{France}, J.~L. {Linsky}, P.~J.~D. {Mauas} and M.~C. {Vieytes},
  \emph{{High stellar FUV/NUV ratio and oxygen contents in the atmospheres of
  potentially habitable planets}},
  \href{https://doi.org/10.1016/j.epsl.2013.10.024}{\emph{Earth Planet. Sci.
  Lett.} {\bfseries 385} (2014) 22--27},
  [\href{https://arxiv.org/abs/1310.2590}{{\ttfamily 1310.2590}}].

\bibitem{KS02}
J.~F. {Kasting} and J.~L. {Siefert}, \emph{{Life and the Evolution of Earth's
  Atmosphere}}, \href{https://doi.org/10.1126/science.1071184}{\emph{Science}
  {\bfseries 296} (2002) 1066--1068}.

\bibitem{Liv99}
M.~{Livio}, \emph{{How Rare Are Extraterrestrial Civilizations, and When Did
  They Emerge?}}, \href{https://doi.org/10.1086/306668}{\emph{Astrophys. J.}
  {\bfseries 511} (1999) 429--431},
  [\href{https://arxiv.org/abs/astro-ph/9808237}{{\ttfamily
  astro-ph/9808237}}].

\bibitem{LFA13}
J.~L. {Linsky}, K.~{France} and T.~{Ayres}, \emph{{Computing Intrinsic
  LY{$\alpha$} Fluxes of F5 V to M5 V Stars}},
  \href{https://doi.org/10.1088/0004-637X/766/2/69}{\emph{Astrophys. J.}
  {\bfseries 766} (2013) 69},
  [\href{https://arxiv.org/abs/1301.5711}{{\ttfamily 1301.5711}}].

\bibitem{LS93}
W.~{Landsman} and T.~{Simon}, \emph{{A catalog of stellar Lyman-alpha fluxes}},
  \href{https://doi.org/10.1086/172589}{\emph{Astrophys. J.} {\bfseries 408}
  (1993) 305--322}.

\bibitem{LFF14}
J.~L. {Linsky}, J.~{Fontenla} and K.~{France}, \emph{{The Intrinsic Extreme
  Ultraviolet Fluxes of F5 V TO M5 V Stars}},
  \href{https://doi.org/10.1088/0004-637X/780/1/61}{\emph{Astrophys. J.}
  {\bfseries 780} (2014) 61},
  [\href{https://arxiv.org/abs/1310.1360}{{\ttfamily 1310.1360}}].

\bibitem{OS12}
L.~M. {Och} and G.~A. {Shields-Zhou}, \emph{{The Neoproterozoic oxygenation
  event: Environmental perturbations and biogeochemical cycling}},
  \href{https://doi.org/10.1016/j.earscirev.2011.09.004}{\emph{Earth Sci. Rev.}
  {\bfseries 110} (2012) 26--57}.

\bibitem{PLK11}
L.~W. {Parfrey}, D.~J.~G. {Lahr}, A.~H. {Knoll} and L.~A. {Katz},
  \emph{{Estimating the timing of early eukaryotic diversification with
  multigene molecular clocks}},
  \href{https://doi.org/10.1073/pnas.1110633108}{\emph{Proc. Natl. Acad. Sci.
  USA} {\bfseries 108} (2011) 13624--13629}.

\bibitem{McK96}
C.~P. {McKay}, \emph{{Time For Intelligence On Other Planets}},  in
  \emph{Circumstellar Habitable Zones} (L.~R. {Doyle}, ed.), pp.~405--419,
  Travis House Publications, 1996.

\bibitem{HA14}
R.~{Heller} and J.~{Armstrong}, \emph{{Superhabitable Worlds}},
  \href{https://doi.org/10.1089/ast.2013.1088}{\emph{Astrobiology} {\bfseries
  14} (2014) 50--66}, [\href{https://arxiv.org/abs/1401.2392}{{\ttfamily
  1401.2392}}].

\bibitem{LiL18}
M.~{Lingam} and A.~{Loeb}, \emph{{Optimal Target Stars in the Search for
  Life}}, \href{https://doi.org/10.3847/2041-8213/aabd86}{\emph{Astrophys. J.
  Lett.} {\bfseries 857} (2018) L17},
  [\href{https://arxiv.org/abs/1803.07570}{{\ttfamily 1803.07570}}].

\bibitem{LL18}
M.~{Lingam} and A.~{Loeb}, \emph{{Role of stellar physics in regulating the
  critical steps for life}}, {\emph{Astrobiology} (2018) },
  [\href{https://arxiv.org/abs/1804.02271}{{\ttfamily 1804.02271}}].

\end{thebibliography}

\providecommand{\href}[2]{#2}\begingroup\raggedright\endgroup

\end{document}